\documentclass[12pt,a4paper,english,amssymb,nofootinbib,superscriptaddress]{revtex4-2}
\usepackage{lmodern}
\usepackage{lmodern}

\usepackage[T1]{fontenc}
\usepackage[latin9]{inputenc}
\usepackage{babel}
\usepackage{fancybox}
\usepackage{calc}
\usepackage{amsmath}
\usepackage{amssymb}
\usepackage{esint}
\usepackage[pdfusetitle,
 bookmarks=true,bookmarksnumbered=false,bookmarksopen=false,
 breaklinks=false,pdfborder={0 0 1},backref=false,colorlinks=false]
 {hyperref}

\makeatletter

\usepackage{babel}

\def\be{\begin{equation}}
\def\ee{\end{equation}}

\@ifundefined{textcolor}{}{%
	\definecolor{BLACK}{gray}{0}
	\definecolor{WHITE}{gray}{1}
	\definecolor{RED}{rgb}{1,0,0}
	\definecolor{GREEN}{rgb}{0,1,0}
	\definecolor{BLUE}{rgb}{0,0,1}
	\definecolor{CYAN}{cmyk}{1,0,0,0}
	\definecolor{MAGENTA}{cmyk}{0,1,0,0}
	\definecolor{YELLOW}{cmyk}{0,0,1,0}
}

\usepackage{latexsym}\usepackage{bm}

\makeatother

\begin{document}
\title{Consistency Problems of Conformal Killing Gravity }
\author{Emel Altas}\footnote{Present address: Department of Engineering Science, Abdullah
Gul University, 38080 Kayseri, Turkey.}
\email{emel.altas@agu.edu.tr}

\affiliation{Department of Physics,~\\
 Karamanoglu Mehmetbey University, 70100, Karaman, Turkey}
\author{Bayram Tekin\thanks{Corresponding author} }
\email{btekin@metu.edu.tr}

\affiliation{Department of Physics,~\\
 Middle East Technical University, 06800, Ankara, Turkey}
\date{\today}
\begin{abstract}
\noindent We show that gravity field equations based on a tensor with rank greater than 2 have consistency problems in the sense that integration constants in the solutions, such as the parameter $m$ in the Schwarzschild metric, do not allow for an interpretation in terms of conserved quantities in the theory.
 The recently introduced Conformal Killing Gravity, an interesting extension of General Relativity that inherits all the solutions of the latter, and is defined with a rank-3 tensor field equation that does not arise from a diffeomorphism-invariant action, is plagued with this problem. In this theory, it is not clear at all how one can define the energy and angular momentum for black hole solutions, or define the analogues of the formulas, such as the quadrupole formula, in the weak field limit for gravitational waves emitted by compact sources.  
\end{abstract}
\maketitle

\section{Introduction}

It is no secret now, due to the large and small scale problems of General Relativity, that some of us wake up in the morning to find a viable modified theory of gravity. In this noble matter, there are some guidelines to be followed. Here, we will point to some of those of which all modified-gravity seekers should be aware. For example, one should be able to answer how one can define, at least in the weak field regime, the quadrupole formula, which requires some approximate definition of the energy-momentum content of gravitational waves. This becomes highly tricky if the source part of the gravity field equations is not the energy-momentum tensor but the derivative of it. 

The second issue is this: Integration constants that appear in the black hole solutions should be tied to the conserved quantities, such as the mass, linear, and angular momentum, or some extra hairs. Pure geometry does not determine the physical meaning of these integration constants in the metric. For example, the Kerr-AdS black hole can be a solution to two distinct gravity theories, yet it might have different energy and angular momentum in these theories, as the coupling constants of the theory conspire and change the conserved charges. This was stressed in \cite{dt1,dt2} in the context of higher-curvature theories of gravity. If this viewpoint is dismissed, then one would have problems in black hole thermodynamics, especially the first law, where variations of the conserved charges are tied to each other.    

A third issue, no less important, is related to the existence of diffeomorphism-invariant {\it action} that yields the field equations of the theory.  Note that it is not the diffeomorphism invariance of the {\it field equations}, that is almost a trivial matter once the field equations are defined via tensors.  Let us expound upon this here and then demonstrate on the example of Conformal Killing Gravity introduced in \cite{Harada}. Take the following matter-free field equations (we take the vacuum field equations for brevity; the arguments are valid when matter is added),
\begin{equation}
\mathcal{H}(g) =0, \label{1.denklem}
\end{equation}
where ${\mathcal{H}}(g)$ can be of some generic rank tensor which is a function of the metric tensor and its derivatives. Here the important point is this:  ${\mathcal{H}}(g)$ is a geometric object such as $\mathcal{H}(g)= H_{\mu \nu \sigma} dx^\mu \otimes dx^\nu \otimes dx^\sigma$. Let $\lambda\in\mathbb{R}$ and $\varphi$
be a one-parameter family of diffeomorphisms on the spacetime manifold (assumed to be pseudo-Riemannian) 
$\varphi:\mathbb{R}\times{\mathcal{M}}\rightarrow{\mathcal{M}}$,
then diffeomorphism invariance of any tensor field ${\mathcal{T}}$ simply boils down to the equality
\begin{equation}
{\mathcal{T}}(\varphi^{*}g)=\varphi^{*}{\mathcal{T}}(g),\label{diffeomorphisminvariance}
\end{equation}
where $\varphi^{*}$ is the pullback of the diffeomorphism.  
This statement is true even for discrete diffeomorphisms, but for a continuous one-parameter family of diffeomorphisms, we can explore further the restrictions coming from this statement.  One must note the role played by the metric tensor in the equation. To proceed, let  $\varphi_{\lambda}$ denote the one-parameter diffeomorphisms, (i.e. $\varphi_\lambda:{\mathcal{M}}\rightarrow{\mathcal{M}}$) with $\varphi_{0}$ defined as the identity map.  Then from (\ref{diffeomorphisminvariance}), one finds 
\begin{equation}
\frac{d}{d\lambda}{\mathcal{T}}(\varphi_\lambda^{*}g)=\frac{d}{d\lambda}\varphi_\lambda^{*}{\mathcal{T}}(g),\label{turev}
\end{equation}
which, evaluated at $\lambda=0$, yields\footnote{For a more compact notation, one can rewrite (\ref{turev}) using the chain rule as \begin{equation}
D{\mathcal{T}}(\varphi_{\lambda}^{*}g)\cdot\frac{d}{d\lambda}\varphi_{\lambda}^{*}g=\varphi_{\lambda}^{*}\left(\mathcal{L}_{X}{\mathcal{T}}(g)\right),\nonumber
\end{equation}
where $D$ denotes the Fr\'echet derivative and $\mathcal{L}_{X}$ denotes the Lie derivative along the vector field $X$. See \cite{Altas_Taub} for this and the relevant references therein, as well as the second derivative of the equation that brings constraints on perturbation theory, so-called the Taub-charges, about a given background solution. }
\begin{equation}
\Bigg[\left.\frac{d}{d\lambda}\right\vert_{\lambda=0} {\mathcal{T}}(\varphi_\lambda^{*}g) \Bigg] =\Bigg[\left.\frac{d}{d\lambda}\right\vert_{\lambda=0}\varphi_\lambda^{*}{\mathcal{T}}(g)\Bigg]. \label{turev2}
\end{equation}
Let us now assume that the one-parameter family of diffeomorphisms is generated by the vector field $X$ and consider our tensor to be the metric tensor, namely, we are looking for the consequences of   (\ref{turev2}) on the metric tensor. In components, it is easy to see that, from (\ref{turev2}), one has
\begin{equation}
(\delta g)_{ \mu \nu} = \nabla_\mu X_\nu +\nabla_\nu X_\mu,
\end{equation}
which, of course, is no surprise, but we must keep in mind that this change in the components of the metric tensor field followed from the diffeomorphism invariance of the metric tensor as a geometric object.   Now, if we apply this procedure verbatim to the Riemann curvature tensor, namely, demanding  
\begin{equation}
\text{Riem}(\varphi^{*}g)=\varphi^{*}\text{Riem}(g),
\end{equation}
one ends up with the Bianchi identities \cite{Kazdan}
\begin{equation}
R_{\mu \nu\sigma \rho}+R_{\mu \sigma \rho \nu}+R_{\mu \rho\nu \sigma }=0, \hskip 1 cm  \nabla_\alpha R_{\mu \nu\sigma \rho}+ \nabla_\sigma R_{\mu \nu \rho\alpha}+ \nabla_\rho R_{\mu \nu\alpha\sigma}=0. \label{Bianchi}
\end{equation}
So, diffeomorphism invariance of the Riemann tensor is not a tautological statement, it leads to these Bianchi identities. From the differential Bianchi identity, or the diffeomorphism invariance of the Ricci curvature, $\text{Ricci}(\varphi^{*}g)=\varphi^{*}\text{Ricci}(g)$,
 the contracted Bianchi identity follows: $\nabla_\mu G^{\mu \nu}=0$ with $G^{\mu \nu}:= R^{\mu \nu} -\frac{1}{2} g^{\mu \nu} R $  \cite{DeTurck}.  The importance of this divergence-free tensor for physics needs no discussion, but we must amplify one important point, as it will appear below for a generic theory of gravity. Divergence-free property of the Einstein tensor either comes from the diffeomorphism invariance of the Ricci tensor, as stated above, or from the diffeomorphism invariance of the action  $\int d^n x \sqrt{-g} R$.  Either way, one finds the same equation,  $\nabla_\mu G^{\mu \nu}=0$,  for {\it all metrics}, not just the ones that are solutions to the Einstein field equations. From the vantage point of physics, $\nabla_\mu G^{\mu \nu}=0$ is linked to the covariant conservation of the matter energy-momentum tensor $\nabla_\mu T^{\mu \nu}=0$, but from the vantage point of mathematics, $\nabla_\mu G^{\mu \nu}=0$ can be read as the following: given a Ricci tensor, one may not be able to find a metric that yields that tensor \cite{DeTurck} if the Einstein tensor is not divergence-free. 
 
 What happens when a theory, defined by some tensorial field equations as (\ref{1.denklem}) lacks a proper action formulation?  Then one can try to find out if there are some non-trivial consequences of the statement that the field equations are tensorial. 
 Typically in these theories, there are some {\it absolute structures}. For example 
 (see \cite{Straumann} and the relevant references therein) if one defines the field equations as the vanishing of the Riemann tensor (which do not arise from the variation of an action with respect to the metric field $g$) 
 $R_{\mu \nu \sigma \rho}(g)=0$, one, obviously, has a diffeomorphism invariant theory, but there is an absolute structure in the theory, which is just the flat space metric $\eta$, and all solutions of the theory are Riemann flat with $g = \varphi^* \eta$, and there are no dynamical degrees of freedom as far as gravity is concerned: This is just Special Relativity cast in the form of tensor field equations and curvature. The invariance group of the theory, that is, the group that leaves the absolute structure invariant, is the Poincar\'e group $ISO(1,n-1)$, which is not the entire diffeomorphism group $\text{Diff}(\mathbb{R}^n)$ as in the case of General Relativity.  So the crucial point here is this: diffeomorphism invariance at the level of the field equations and the level of an action can lead to entirely different theories. General Relativity is an exception in this sense. 
 
To make the preceding discussion concrete, we now consider the particular gravity theory defined in \cite{Harada} to explain the large-distance behavior of gravity without dark energy. 
 
 \section{Conformal Killing Gravity}
 We shall work in four dimensions and in a coordinate-adapted chart. The suggested field equations \cite{Harada} are\footnote{Let us note that an equivalent form of these equations was given in \cite{Mantica}, and we keep their notation 
\begin{equation}
R_{jk} - \frac{R}{2} g_{jk} = T_{jk} + K_{jk}\nonumber 
\end{equation}
where \( K_{jk} \) is a divergence-free conformal Killing tensor of which the explicit form is 
\begin{equation}
\nabla_j K_{kl} + \nabla_k K_{lj} + \nabla_l K_{jk} = \frac{1}{6} 
\left( g_{kl} \nabla_j K + g_{lj} \nabla_k K + g_{jk} \nabla_l K \right).\nonumber
\end{equation}
In this formulation, the existence of a divergence-free conformal Killing tensor seems to be the absolute structure described above. 
}
\begin{equation}
H_{\mu\nu\sigma}=8\pi G_N T_{\mu\nu\sigma},\label{field=000020equations}
\end{equation}
where the $H$-tensor is defined as 
\begin{align}
H_{\mu\nu\sigma}:= & \nabla_{\mu}R_{\nu\sigma}+\nabla_{\nu}R_{\mu\sigma}+\nabla_{\sigma}R_{\mu\nu}-\frac{1}{3}\left(g_{\nu\sigma}\nabla_{\mu}R+g_{\mu\sigma}\nabla_{\nu}R+g_{\mu\nu}\nabla_{\sigma}R\right),\label{definitionH}
\end{align}
while the source term reads
\begin{align}
T_{\mu\nu\sigma}:=\nabla_{\mu} & T_{\nu\sigma}+\nabla_{\nu}T_{\mu\sigma}+\nabla_{\sigma}T_{\mu\nu}-\frac{1}{6}\left(g_{\nu\sigma}\nabla_{\mu}T+g_{\mu\sigma}\nabla_{\nu}T+g_{\mu\nu}\nabla_{\sigma}T\right). \label{3index}
\end{align}
Here $T_{\nu\sigma}$ is the usual covariantly-conserved energy-momentum tensor, so $T_{\mu\nu\sigma}$ is not directly the energy-momentum tensor, which will be one of the problems in trying to construct, say, the energy-momentum of gravitational waves, or black holes in this theory. $H_{\mu\nu\sigma}$ and  $T_{\mu\nu\sigma}$ are totally symmetric and traceless. The tracelessness of $H_{\mu\nu\sigma}$ follows from the contracted Bianchi identity, while the tracelessness of $T_{\mu\nu\sigma}$ follows from covariant conservation of the energy-momentum tensor.  In 4 dimensions, there are 16 linearly independent equations in this theory, 6 more than in General Relativity.

 It is clear that with 3 indices, the $H$-tensor cannot come from an action, with the metric being the dynamical field. Hence, as per the above discussion, the theory has an important problem: it lacks the generalized Bianchi identities for all metrics.
In fact, as we shall show, for the vacuum case,  $\nabla^\mu H_{\mu\nu\sigma}$,  $\nabla^\mu \nabla^\nu H_{\mu\nu\sigma}$ and $\nabla^\mu \nabla^\nu \nabla^\sigma H_{\mu\nu\sigma}$  vanish only on-shell, namely only for the solutions of the field equations. So, these on-shell Bianchi identities do not arise from the diffeomorphism invariance of the theory. Of course, the reader might wonder what, if any, conditions arise from the fact that we have diffeomorphism invariance of the $H$-tensor itself?  So one needs to compute under what conditions one has
\begin{equation}
\Bigg[\left.\frac{d}{d\lambda}\right\vert_{\lambda=0} H(\varphi_\lambda^{*}g) \Bigg]_{\mu \nu \sigma} =\Bigg[\left.\frac{d}{d\lambda}\right\vert_{\lambda=0}\varphi_\lambda^{*}H(g)\Bigg]_{\mu \nu \sigma}. \label{H-turev}
\end{equation}
This is a somewhat lengthy computation, but the astute reader can already guess the result: this equation will be satisfied as long as the usual Bianchi identities on the Riemann tensor (\ref{Bianchi}) are satisfied, there is no constraint on the $H$-tensor. This, in some sense, is the beauty of Riemannian geometry: All tensors based on the metric tensor, as geometric objects, are diffeomorphism invariant; the only conditions are the algebraic and differential Bianchi identities on the Riemann tensor. 

\subsection{On-shell Bianchi identities on the $H$-tensor}

Let us consider the vacuum case, set $H^{\mu\nu\sigma}=0$ and, assuming this, find the consequences (integrability conditions). Einstein manifolds are clearly solutions of this theory, but there are also non-Einsteinian solutions. 
(In this work, we are not going to discuss solutions of the theory; the interested reader can check \cite{sol1,sol2,sol3,sol4,sol5,sol6,sol7}.) Clearly, we must have $\nabla_\mu H^{\mu\nu\sigma}=0$, which is correct as long as $H^{\mu\nu\sigma}=0$. This readily follows in a local inertial frame (LIF) for which $\nabla_\mu H^{\mu\nu\sigma} = \partial _{\hat\mu} H^{\hat \mu \hat \nu \hat \sigma}=0$, assuming $H^{\hat \mu \hat \nu \hat \sigma}=0$, where the hat denotes the LIF coordinates with $\Gamma^{\hat\mu}_{\hat \nu \hat \sigma}=0$. So, the theory has an on-shell covariant conservation or a generalized Bianchi identity. But this is almost a trivial statement: As long as the metric is sufficiently differentiable, any theory will have this. The crucial point is that the covariant divergence of the tensor is not automatically zero for {\it all} metrics. Let us show this as it is, by no means clear. Taking  $H$  from (\ref{definitionH}), which is in dire need of differentiation. So  we have
\begin{equation}
\begin{aligned}\nabla_{\mu}H^{\mu\nu\sigma}= & \nabla_{\mu}\nabla^{\mu}R^{\nu\sigma}+\nabla_{\mu}\nabla^{\nu}R^{\mu\sigma}+\nabla_{\mu}\nabla^{\sigma}R^{\mu\nu}-\frac{1}{3}\left(g^{\nu\sigma}\nabla_{\mu}\nabla^{\mu}R+\nabla^{\sigma}\nabla^{\nu}R+\nabla^{\nu}\nabla^{\sigma}R\right),\end{aligned}
\end{equation}
which yields
\begin{equation}
\begin{aligned}\nabla_{\mu}H^{\mu\nu\sigma}=\square R^{\nu\sigma} & +\nabla_{\mu}\nabla^{\nu}R^{\mu\sigma}+\nabla_{\mu}\nabla^{\sigma}R^{\mu\nu}-\frac{1}{3}g^{\nu\sigma}\square R-\frac{2}{3}\nabla^{\sigma}\nabla^{\nu}R,\end{aligned}
\end{equation}
where $\square=\nabla_{\sigma}\nabla^{\sigma}$. The divergence can
be rewritten as
\begin{equation}
\begin{aligned}\nabla_{\mu}H^{\mu\nu\sigma} & = & \square R^{\nu\sigma}+\left[\nabla_{\mu},\nabla^{\nu}\right]R^{\mu\sigma}+\nabla^{\nu}\nabla_{\mu}R^{\mu\sigma}+\left[\nabla_{\mu},\nabla^{\sigma}\right]R^{\mu\nu}\\
 &  & +\nabla^{\sigma}\nabla_{\mu}R^{\mu\nu}-\frac{1}{3}g^{\nu\sigma}\square R-\frac{2}{3}\nabla^{\sigma}\nabla^{\nu}R.
\end{aligned}
\end{equation}
Using the contracted Bianchi identity, $\nabla_{\sigma}R^{\sigma\mu}=\frac{1}{2}\nabla^{\mu}R$,
and the Ricci identity
\begin{equation}
\left[\nabla_{\mu},\nabla_{\nu}\right]X_{\sigma}=R_{\mu\nu\sigma\lambda}X^{\lambda},
\end{equation}
one arrives at
\begin{align}
\nabla_{\mu}H^{\mu\nu\sigma}=\square R^{\nu\sigma} & +2R_{\lambda}^{\nu}R^{\lambda\sigma}+2 R_{\mu\lambda}R^{\mu\nu\sigma\lambda}-\frac{1}{3}g^{\nu\sigma}\square R+\frac{1}{3}\nabla^{\sigma}\nabla^{\nu}R,\label{firstdivergence}
\end{align}
which does not vanish for all metrics. This was our first point. But, it does vanish for the solutions of the theory as one can see with the following computation: since $H_{\mu\nu\sigma}=0$, one has 
\begin{equation}
\begin{aligned}\nabla_{\mu}H^{\mu\nu\sigma}= & -\nabla_{\mu}\nabla^{\nu}R^{\mu\sigma}-\nabla_{\mu}\nabla^{\sigma}R^{\mu\nu}+\nabla^{\sigma}\nabla^{\nu}R+2R^{\nu\lambda}R_{\lambda}^{\sigma}+2R_{\mu\lambda}R^{\mu\nu\sigma\lambda}.\end{aligned}
\end{equation}
Then we have
\begin{equation}
\begin{gathered}\nabla_{\mu}H^{\mu\nu\sigma}=-\left[\nabla_{\mu},\nabla^{\nu}\right]R^{\mu\sigma}-\nabla^{\nu}\nabla_{\mu}R^{\mu\sigma}-\left[\nabla_{\mu},\nabla^{\sigma}\right]R^{\mu\nu}-\nabla^{\sigma}\nabla_{\mu}R^{\mu\nu}\\
+\nabla^{\sigma}\nabla^{\nu}R+2R^{\nu\lambda}R_{\lambda}^{\sigma}+2R_{\mu\lambda}R^{\mu\nu\sigma\lambda},
\end{gathered}
\end{equation}
which yields
\begin{equation}
\begin{aligned}\nabla_{\mu}H^{\mu\nu\sigma}= & -R_{\mu}{}^{\nu\mu}\ _{\lambda}R^{\lambda\sigma}-R_{\mu}{}^{\nu\sigma}\ _{\lambda}R^{\mu\lambda}-\frac{1}{2}\nabla^{\nu}\nabla^{\sigma}R-R_{\mu}{}^{\sigma\mu}{}_{\lambda}R^{\lambda\nu}-R_{\mu}\ ^{\sigma\nu}{}_{\lambda}R^{\mu\lambda}\\
 & -\frac{1}{2}\nabla^{\sigma}\nabla^{\nu}R+\nabla^{\sigma}\nabla^{\nu}R+2R^{\nu\lambda}R_{\lambda}^{\sigma}+2R_{\mu\lambda}R^{\mu\nu\sigma\lambda},
\end{aligned}
\end{equation}
and reduces to
\begin{equation}
\begin{aligned}\nabla_{\mu}H^{\mu\nu\sigma}= & -R^{\nu\lambda}R_{\lambda}^{\sigma}-R^{\mu\nu\sigma\lambda}R_{\mu\lambda}-\nabla^{\nu}\nabla^{\sigma}R-R^{\sigma\lambda}R_{\lambda}^{\nu}-R^{\mu\sigma\nu\lambda}R_{\mu\lambda}\\
 & +\nabla^{\sigma}\nabla^{\nu}R+2R^{\nu\lambda}R_{\lambda}^{\sigma}+2R_{\mu\lambda}R^{\mu\nu\sigma\lambda},
\end{aligned}
\end{equation}
where all the terms on the right-hand side cancel each other as expected from the left-hand side of the equation.
We can extend this computation to the second and third derivatives of the $H$-tensor and show that they do not vanish automatically for all metrics but only for solutions. In fact, in the next section, we shall show the explicit expression of the second derivative as we need it for the construction of a conserved current. Before we start that discussion, let us note in passing that the following form of the field equations might be useful in certain computations, such as finding solutions: Defining
\begin{equation}
\tilde{R}_{\mu\nu}:=R_{\mu\nu}-\frac{1}{3}g_{\mu\nu}R,\label{R_tilda}
\end{equation}
one has
\begin{equation}
H_{\mu\nu\sigma}=\nabla_{\mu}\tilde{R}_{\nu\sigma}+\nabla_{\nu}\tilde{R}_{\mu\sigma}+\nabla_{\sigma}\tilde{R}_{\mu\nu},
\end{equation}
and introducing the constant tensor
\begin{equation}
{\mathcal{I}}_{\mu\nu\sigma}^{\alpha\beta\rho}:=\frac{1}{3}\left(\delta_{\mu}^{\alpha}\delta_{(\nu}^{\beta}\delta_{\sigma)}^{\rho}+\delta_{\nu}^{\alpha}\delta_{(\mu}^{\beta}\delta_{\sigma)}^{\rho}+\delta_{\sigma}^{\alpha}\delta_{(\mu}^{\beta}\delta_{\nu)}^{\rho}\right),
\end{equation}
then one has 
\begin{equation}
H_{\mu\nu\sigma}=3
{\mathcal{I}}_{\mu\nu\sigma}^{\alpha\beta\rho}\ \nabla_{\alpha}\tilde{R}_{\beta\rho}, \hskip 1 cm 
{\mathcal{I}}_{\mu\nu\sigma}^{\alpha\beta\rho}\ H_{\alpha\beta\rho}=H_{\mu\nu\sigma}.\label{contraction1}
\end{equation}
The field equations read 
\begin{equation}
3{\mathcal{I}}_{\mu\nu\sigma}^{\alpha\beta\rho}\ \nabla_{\alpha}\tilde{R}_{\beta\rho}=8\pi GT_{\mu\nu\sigma}.\label{compactfieldeq}
\end{equation}

\subsection{Construction of the Conserved Charges}

To be able to identify the arbitrary integration constants in the solutions of the field equations with some physical quantities, such as the mass-energy, angular momentum of black holes, or some other physical objects, we need to find the conserved charges in this theory.  There are two separate obstacles to this theory. The fact that one does not have a diffeomorphism-invariant action makes this rather difficult; secondly, the field equations on the right-hand side involve the derivatives of the energy-momentum tensor, not the energy-momentum tensor itself, which hampers the identification of gravitational energy. 

If one had a diffeomorphism invariant action, one would crank up the Noether procedure \cite{Banados}:  Say $S[g]$ is the action, look at the symmetry variations $\delta_s S[g, \delta_s g]$ which should yield, by definition of a symmetry transformation, only a divergence (a boundary term) as
\begin{equation}
 \delta_s S[g, \delta_s g] = \int_{\mathcal{M}} d^n x \,\partial_\mu K^\mu, \hskip 1 cm  \forall\,\, g,
\end{equation}
and compute  generic variations
\begin{equation}
 \delta S[g, \delta g] = \int_{\mathcal{M}} d^n x \,{\mathcal{E}} \delta g +  \int d^n x \,\partial_\mu \Phi^\mu, \hskip 1 cm  \forall \,\, \delta g.
\end{equation}
These two expressions will be equal to each other, if one considers only the solutions of the theory $g = \bar g$, for which the Euler-Lagrange tensor vanishes: ${\mathcal{E}}=0$, and one sets $\delta g =  \delta_s g$ in the second equation, while setting $g= \bar g$ in the first one, which yields a partially conserved current
$ J^\mu = K^\mu - \Phi^\mu,$ up to a divergence-free vector. 

The absence of an action-formulation does not allow us to use this powerful procedure. But we can still try to emulate what is usually done for General Relativity and higher curvature theories  (see \cite{Sisman}) that still has the usual energy-momentum tensor on the right-hand side of the field equations.  Briefly the procedure is as follows: given some field equations of the form
\begin{equation}
\bar{\mathcal{E}}_{\mu \nu}(g, \nabla \text{Riem},...) = \kappa \tau_{\mu \nu},\label{exact}
\end{equation}
with the Bianchi identity $\nabla^\mu \mathcal{E}_{\mu \nu}(g, \nabla \text{Riem},...) =0$, one defines a vacuum solution ($\tau_{\mu \nu}=0$), the background, $\bar{g}$ with vanishing conserved charges, that has at least one, but usually more, Killing vectors ($\bar{\xi}_a^\mu$). Then for any other solution, $g_{\mu \nu}$ that asymptotically has the same symmetry as the background, and is related to the background solution via 
$g_{\mu \nu}= \bar{g}_{\mu \nu} + \kappa h_{\mu \nu}$, where $h_{\mu\nu}$ need not be small everywhere in spacetime, but should decay sufficiently fast at infinity, one splits (\ref{exact}) as 
\begin{eqnarray}
\kappa {\cal {O}}(\bar{g})_{\mu\nu\alpha\beta}h^{\alpha\beta}=\kappa T_{\mu\nu} (\tau, \kappa h^2, \kappa^2 h^3,...),\label{ope-1}
\end{eqnarray}
where we used the vanishing of the background equation. Observe that the matter content and all the non-linear terms in the gravitational field are on the right-hand side. The background operator ${\cal {O}}(\bar{g})$ depends on the particular theory at hand and can be rather complicated in generic gravity theories. But, in GR with a cosmological constant, one has 
${\cal G}_{\mu\nu}^{(1)}=:{\cal O}\left(\bar{g}\right)_{\mu\nu\alpha\beta}  h^{\alpha \beta}$ with \cite{Tekin_trib}
\begin{align}
{\cal O}\left(\bar{g}\right)_{\mu\nu\alpha\beta}:= & \frac{1}{2}\bar{g}_{\mu\nu}\bar{g}_{\alpha\beta}\left(\bar{\Box}+\frac{2\Lambda}{n-2}\right)-\frac{1}{2}\bar{g}_{\mu\alpha}\bar{g}_{\nu\beta}\left(\bar{\Box}+\frac{4\Lambda}{n-2}\right)-\frac{1}{2}\left(\bar{g}_{\mu\nu}\bar{\nabla}_{\alpha}\bar{\nabla}_{\beta}+\bar{g}_{\alpha\beta}\bar{\nabla}_{\mu}\bar{\nabla}_{\nu}\right)\nonumber \\
 & +\frac{1}{2}\left(\bar{g}_{\mu\alpha}\bar{\nabla}_{\beta}\bar{\nabla}_{\nu}+\bar{g}_{\nu\beta}\bar{\nabla}_{\alpha}\bar{\nabla}_{\mu}\right),
\end{align}
and in a particular gauge, one can find the Green's function which is the inverse of this operator to relate the field $h_{\mu \nu}$ to the source $T_{\mu \nu}$ at least in perturbation theory. The quadrupole formula relating the third derivative of the traceless quadrupole moment of a weakly gravitating system to the power loss resulting from the gravitational waves follows from this in the lowest order. 

One crucial piece of information is the following: the exact Bianchi identity $\nabla_\mu {\mathcal{E}}^{\mu \nu}=0$ yields a linearized on-shell Bianchi identity 
$\bar{\nabla}_\mu {\mathcal{E}}_{(1)}^{\mu \nu}=0$ where ${\mathcal{E}}_{(1)}^{\mu \nu} :=  {\cal {O}}(\bar{g})_{\mu\nu\alpha\beta}h^{\alpha\beta}= T_{\mu\nu} (\tau, \kappa h^2, \kappa^2 h^3,...)$. Then the following current
\begin{equation}
\mathcal{J}^\mu  := \sqrt{-\bar{g}}\,{\mathcal{E}}_{(1)}^{\mu \nu}\,\bar{\xi}_{\nu}
\label{suslu}
\end{equation}
is conserved, $\partial_\mu \mathcal{J}^\mu=0$, and leads to the conserved charge 
\begin{equation}
Q[\bar \xi, \bar{\Sigma}_t ]: =\int_{\bar{\Sigma}_t } d^{n-1}y\,
\sqrt{\bar{\gamma}}\, {\hat n}_\mu \, \bar{\xi}_{\nu}\,{\mathcal{E}}_{(1)}^{\mu \nu},
\label{charge1}
\end{equation}
where $\gamma_{\mu \nu}$ is the pullback metric on $\bar{\Sigma}_t$ which is a spatial hypersurface at any time $t$.  Furthermore, one can find an anti-symmetric tensor  $\mathcal{F}^{\mu \nu}(\bar \xi, h)$, such that  $\bar{\xi}_{\nu}{\mathcal{E}}_{(1)}^{\mu \nu} =: \bar \nabla_\nu  \mathcal{F}^{\mu \nu}(\bar \xi, h)$, which reduces the charge formula to a boundary integral at spatial infinity
\begin{equation}
Q[\bar \xi,\bar{\Sigma}_t ] =\int_{\bar{\Sigma}_t} d^{n-1}y\
\sqrt{\bar{\gamma}}\, {\hat n}_\mu \bar \nabla_\nu  \mathcal{F}^{\mu \nu}(\bar \xi, h) = \int_{\partial \bar{\Sigma}_t} d^{n-2}z\
\sqrt{\bar{q}}\, {\hat n}_\mu \hat \sigma_\nu  \mathcal{F}^{\mu \nu}(\bar \xi, h), 
\label{charge2}
\end{equation}
where $\bar q_{\mu \nu}$ is the induced metric on $\partial \bar{\Sigma}$; and  $\hat \sigma_\nu$ is the outward unit normal co-vector on it. Given gravity theory, one can determine what $\mathcal{F}^{\mu \nu}(\bar \xi, h)$ is. However, for the theory at hand (\ref{field=000020equations}), it is not clear at all which tensor or pseudotensor can represent the gravitational part of the energy.  One can certainly carry out the above-described procedure and expand the field equations about a background solution as
\begin{equation}
\kappa H_{(1)}^{\mu\nu\sigma}  =\kappa T^{\mu\nu\sigma} + \sum_{i=2}^\infty \kappa^i H_{(i)}^{\mu\nu\sigma}, \label{uzun}
\end{equation}
where we have moved all but the linear terms to the right-hand side and set the background expression to zero by definition ${\bar H}^{\mu\nu\sigma}(\bar 
g)=0$.  So the 3-index object (a background, or a pseudo-tensor)
\begin{equation}
\theta^{\mu\nu\sigma} := T^{\mu\nu\sigma} + \sum_{i=2}^\infty \kappa^{i-1} H_{(i)}^{\mu\nu\sigma},
\end{equation}
is supposed to somehow represent the localized matter energy-momentum plus the gravitational energy. One could envisage writing all the higher-order terms
$ \sum_{i=2}^\infty \kappa^{i-1} H_{(i)}^{\mu\nu\sigma}$  as (\ref{3index})
\begin{equation}
 \sum_{i=2}^\infty \kappa^{i-1} H_{(i)}^{\mu\nu\sigma}=:\bar \nabla_{\mu}  \theta_{\nu\sigma}+\bar\nabla_{\nu}\theta_{\mu\sigma}+\bar\nabla_{\sigma}\theta_{\mu\nu}-\frac{1}{6}\left(\bar g_{\nu\sigma}\bar\nabla_{\mu}\theta+\bar g_{\mu\sigma}\bar\nabla_{\nu}\theta+\bar g_{\mu\nu}\bar\nabla_{\sigma}\theta\right),
\end{equation}
and identify $\theta_{\mu\sigma}$ as the gravitational (non-matter) part of the energy-momentum tensor, but this is highly non-unique, not to mention ad-hoc. Note that in practice, one should use the left-hand side of (\ref{uzun}) to define $\theta_{\mu\sigma}$, as the right-hand side is unwieldy. So then one has the following equation for the  Conformal Killing Gravity, as an analogous equation to (\ref{ope-1})
\begin{eqnarray}
\kappa {\cal {O}}(\bar{g})_{\mu\nu\alpha\beta\sigma }h^{\beta \sigma}=\kappa \theta_{\mu\nu\alpha} (\tau, \kappa h^2, \kappa^2 h^3,...).\label{ope-2}
\end{eqnarray}
Even if one can solve this equation perturbatively, and determine the weak gravitational field $h^{\beta \sigma}$,
it is clear that the gravitational field is determined by the derivative of the energy-momentum tensor and not the energy-momentum tensor itself. The usual quadrupole formula fails. Moreover, in GR, using the field equations, one can assign an average energy-momentum tensor to the gravitational wave which reads (in the transverse traceless gauge around flat spacetime) as 
\begin{equation}
 T_{\mu \nu} =\frac{1}{32 \pi} \Big \langle \partial_\mu h_{\sigma \rho} \,\partial_\nu h^{\sigma \rho} \Big \rangle, 
\end{equation}
where the averaging is done over several wavelengths. We cannot use this formula for the Conformal Killing Gravity as this formula was obtained through the field equations of the theory and in this construction the weak field $h_{\sigma \rho}$ directly is source of second order perturbation theory via the non-averaged form of  $T_{\mu \nu}$.  

\subsubsection{Direct approach}

Due to these obstacles, let us try to get a conserved current by directly manipulating the field equations in their explicit form.  Let us first consider only the matter-free case.
We found above that the first divergence of the $H$-tensor is
\begin{align}
\nabla_{\mu}H^{\mu\nu\sigma}=\square R^{\nu\sigma} & +2R_{\lambda}^{\nu}R^{\lambda\sigma}+2R_{\mu\lambda}R^{\mu\nu\sigma\lambda}-\frac{1}{3}g^{\nu\sigma}\square R+\frac{1}{3}\nabla^{\sigma}\nabla^{\nu}R.\label{firstdivergence}
\end{align}
Taking one more divergence of the $H$-tensor with respect to one
of its remaining two free indices
\begin{equation}
\begin{aligned}\nabla_{\nu}\nabla_{\mu}H^{\mu\nu\sigma}= & \nabla_{\nu}\nabla_{\mu}\nabla^{\mu}R^{\nu\sigma}+2\left(\nabla_{\nu}R_{\lambda}^{\nu}\right)R^{\lambda\sigma}+2R_{\lambda}^{\nu}\nabla_{\nu}R^{\lambda\sigma}\\
 & +2(\nabla_{\nu}R_{\mu\lambda})R^{\mu\nu\sigma\lambda}+2R_{\mu\lambda}\nabla_{\nu}R^{\mu\nu\sigma\lambda}\\
 & -\frac{1}{3}\nabla^{\sigma}\nabla_{\mu}\nabla^{\mu}R+\frac{1}{3}\nabla_{\nu}\nabla^{\sigma}\nabla^{\nu}R,
\end{aligned}
\end{equation}
and using some algebra, one arrives at 
\begin{equation}
\begin{aligned}\nabla_{\nu}\nabla_{\mu}H^{\mu\nu\sigma}= & 4R^{\mu\nu\sigma\lambda}\nabla_{\nu}R_{\mu\lambda}+\frac{7}{3}R^{\lambda\sigma}\nabla_{\lambda}R+6R_{\mu\lambda}\nabla_{\mu}R^{\lambda\sigma}-3R_{\mu\lambda}\nabla^{\sigma}R^{\mu\lambda}+\frac{1}{2}\nabla^{\sigma}\square R.\end{aligned}
\end{equation}
It pays to write the first and second derivatives of the $H$-tensor in terms of the cosmological Einstein tensor 
\begin{equation}
\mathcal{G}_{\mu\nu}=R_{\mu\nu}-\frac{1}{2}g_{\mu\nu}R+\Lambda g_{\mu\nu}.\label{cosmologicalEinstein}
\end{equation}
One has 
\begin{align}
\nabla_{\mu}H^{\mu\nu\sigma}=\square\mathcal{G}^{\nu\sigma} & +2\mathcal{G}_{\lambda}^{\nu}\mathcal{G}^{\lambda\sigma}+2\mathcal{G}_{\mu\lambda}R^{\mu\nu\sigma\lambda}+R\mathcal{G}^{\nu\sigma}-2\Lambda\mathcal{G}^{\nu\sigma}+\frac{1}{6}g^{\nu\sigma}\square R+\frac{1}{3}\nabla^{\sigma}\nabla^{\nu}R,\label{firstdivergenceintermsofEinstein}
\end{align}
and 
\begin{equation}
\begin{aligned}\nabla_{\nu}\nabla_{\mu}H^{\mu\nu\sigma}= & 4R^{\mu\nu\sigma\lambda}\nabla_{\nu}\mathcal{G}_{\mu\lambda}+\frac{10}{3}\mathcal{G}{}^{\lambda\sigma}\nabla_{\lambda}R+6\mathcal{G}_{\mu\lambda}\nabla_{\mu}\mathcal{G}^{\lambda\sigma}-3\mathcal{G}_{\mu\lambda}\nabla^{\sigma}\mathcal{G}^{\mu\lambda}\\
 & -\frac{19\Lambda}{3}\nabla^{\sigma}R+\frac{4}{3}R\nabla^{\sigma}R-\frac{1}{2}\nabla^{\sigma}\square R.
\end{aligned}
\label{seconddivergence}
\end{equation}
 We can rewrite the last equation using
\begin{equation}
\begin{aligned}R^{\mu\nu\sigma\lambda}\nabla_{\nu}\mathcal{G}_{\mu\lambda}  =\nabla_{\nu}\left(\mathcal{G}_{\mu\lambda}R^{\mu\nu\sigma\lambda}\right)-\mathcal{G}_{\mu\lambda}\nabla_{\nu}R^{\mu\nu\sigma\lambda}
  =\nabla_{\nu}\left(\mathcal{G}_{\mu\lambda}R^{\mu\nu\sigma\lambda}\right)-\mathcal{G}_{\mu\lambda}\nabla_{\nu}R^{\nu\mu\lambda\sigma}.
\end{aligned}
\end{equation}
Using the second Bianchi identity one more time, one has
\begin{equation}
\begin{aligned}R^{\mu\nu\sigma\lambda}\nabla_{\nu}\mathcal{G}_{\mu\lambda} & =\nabla_{\nu}\left(\mathcal{G}_{\mu\lambda}R^{\mu\nu\sigma\lambda}\right)-\mathcal{G}_{\mu\lambda}\left(\nabla^{\lambda}R^{\mu\sigma}-\nabla^{\sigma}R^{\mu\lambda}\right)\\
 & =\nabla_{\nu}\left(\mathcal{G}_{\mu\lambda}R^{\mu\nu\sigma\lambda}\right)-\mathcal{G}_{\mu\lambda}\nabla^{\lambda}R^{\mu\sigma}+\mathcal{G}_{\mu\lambda}\nabla^{\sigma}R^{\mu\lambda},
\end{aligned}
\end{equation}
explicitly we have
\begin{equation}
\begin{aligned}R^{\mu\nu\sigma\lambda}\nabla_{\nu}\mathcal{G}_{\mu\lambda}= & \frac{1}{2}\nabla^{\sigma}\left(\mathcal{G}_{\mu\lambda}\mathcal{G}^{\mu\lambda}\right)+\nabla_{\nu}\left(\mathcal{G}_{\mu\lambda}R^{\mu\nu\sigma\lambda}\right)\\
- & \mathcal{G}_{\mu\lambda}\nabla^{\lambda}\mathcal{G}^{\mu\sigma}-\frac{1}{2}\nabla^{\mu}(R\mathcal{G}_{\mu}^{\sigma})+2\Lambda\nabla^{\sigma}R-\frac{1}{4}\nabla^{\sigma}R^{2}.
\end{aligned}
\end{equation}
Finally we have 
\begin{equation}
\begin{aligned}R^{\mu\nu\sigma\lambda}\nabla_{\nu}\mathcal{G}_{\mu\lambda} & =\nabla^{\sigma}\left(\frac{1}{2}\mathcal{G}{}_{\mu\lambda}\mathcal{G}^{\mu\lambda}+2\Lambda R-\frac{1}{4}R^{2}\right)\\
 & +\nabla_{\mu}\left(\mathcal{G}_{\nu\lambda}R^{\nu\mu\sigma\lambda}-\frac{1}{2}R\mathcal{G}^{\mu\sigma}-\mathcal{G}^{\nu\sigma}\mathcal{G}_{\nu}^{\mu}\right).
\end{aligned}
\end{equation}
Inserting these results in equation (\ref{seconddivergence}), we
arrive at the second divergence of the $H$- tensor as
\begin{equation}
\begin{aligned}\nabla_{\nu}\nabla_{\mu}H^{\mu\nu\sigma}= & \nabla^{\sigma}\left(\frac{1}{2}\mathcal{G}{}_{\mu\lambda}\mathcal{G}^{\mu\lambda}+\frac{5}{4}\Lambda R-\frac{1}{3}R^{2}+\frac{1}{2}\square R\right)\\
 & +\nabla_{\mu}\left(4\mathcal{G}_{\nu\lambda}R^{\nu\mu\sigma\lambda}+\frac{4}{3}R\mathcal{G}^{\mu\sigma}+2\mathcal{G}^{\nu\sigma}\mathcal{G}_{\nu}^{\mu}\right).
\end{aligned}
\label{eq:seconddiv}
\end{equation}
As we noted the $H$-tensor is not divergence-free at the first and second
orders off-shell. However, we shall use these expressions to construct a conserved current.

From the right-hand-side of (\ref{eq:seconddiv}), we define symmetric second rank tensor 
\begin{equation}
\begin{aligned}\Phi^{\mu\sigma}=  g^{\mu \sigma}\left(\frac{1}{2}\mathcal{G}{}_{\alpha\beta}\mathcal{G}^{\alpha\beta}+\frac{5}{4}\Lambda R-\frac{1}{3}R^{2}+\frac{1}{2}\square R\right)
 +4\mathcal{G}_{\nu\lambda}R^{\nu\mu\sigma\lambda}+\frac{4}{3}R\mathcal{G}^{\mu\sigma}+2\mathcal{G}^{\nu\sigma}\mathcal{G}_{\nu}^{\mu},
\end{aligned}
\label{Phi}
\end{equation}
which is divergence-free on-shell, since  $\nabla_\mu \Phi^{\mu\sigma} =\nabla_{\nu}\nabla_{\mu}H^{\mu\nu\sigma} =0$. Then given a Killing vector $\xi^\sigma$, we can find a  conserved current (not worrying about the overall dimensional factor):
\begin{equation}
\mathcal{J}^\mu : = \sqrt{-g} \xi_\sigma \Phi^{\mu\sigma}, \hskip 2 cm \partial_\mu \mathcal{J}^\mu =0. \label{current0}
\end{equation}
It looks like, finally, we have found a viable conserved current in the theory. But observe that for all Einstein manifolds, which are solutions to the Conformal Killing Gravity, this current is identically zero; hence, this conserved current cannot be used to define nontrivial conserved charges to Schwarzschild or Kerr black holes, which are solutions to the theory. This is not acceptable since no black hole spacetime cannot be allowed to have the same energy and angular momentum as a black hole spacetime. If it did, creating black holes would cost nothing.    

Note that the above computation can be extended to include the source term in the following way: Assuming $\xi_{\sigma}$ to be a Killing co-vector, the following statement is exact (no field equations are used) due to symmetries
\begin{equation}
\left(\nabla_{\mu}H^{\mu\nu\sigma}\right)\nabla_{\nu}\xi_{\sigma}=0,
\end{equation}
which yields
\begin{equation}
\nabla_{\nu}\left(\xi_{\sigma}\nabla_{\mu}H^{\mu\nu\sigma}\right)-\xi_{\sigma}\nabla_{\nu}\nabla_{\mu}H^{\mu\nu\sigma}=0.\label{identity}
\end{equation}
Inserting (\ref{eq:seconddiv}) in last equation, one has
\begin{equation}
\begin{aligned} & \nabla_{\mu}\left(\xi_{\sigma}\nabla_{\nu}H^{\mu\nu\sigma}-4\xi_{\sigma}\mathcal{G}_{\nu\lambda}R^{\nu\mu\sigma\lambda}-\frac{4}{3}\xi_{\sigma}R\mathcal{G}^{\mu\sigma}-2\xi_{\sigma}\mathcal{G}^{\nu\sigma}\mathcal{G}_{\nu}^{\mu}\right.\\
 & -\frac{1}{2}\xi^{\mu}\mathcal{G}{}_{\sigma\lambda}\mathcal{G}^{\sigma\lambda}-\frac{5}{4}\xi^{\mu}\Lambda R+\frac{1}{3}\xi^{\mu}R^{2}-\frac{1}{2}\xi^{\mu}\square R\biggr)=0,
\end{aligned}
\end{equation}
which leads to an extended version of the  conserved current (\ref{current0})
\begin{equation}
J^{\mu}:=  -\sqrt{-g}\xi_{\sigma}\nabla_{\nu}H^{\mu\nu\sigma} + \mathcal{J}^\mu. \label{current4}
\end{equation}
Again, for Einstein manifolds, it gives identically zero charges in this theory: if a black hole has the same conserved charges as the background (vacuum) in a theory, then clearly the theory is ill-defined as in the Cotton Gravity case \cite{Altas_son}.

\section{CONCLUSIONS}

We showed that a theory based on a third-rank geometric tensor, of which the source is the first derivative of the usual energy-momentum tensor, is hard to interpret as a gravity theory for several reasons. One cannot define a symmetric pseudo-tensor for gravitational waves, or the total mass and angular momentum of a black hole spacetime. The lack of a diffeomorphism invariant action in the formulation of the theory is also a problem: The field equations satisfy generalized Bianchi identities only on-shell. Such theories usually have absolute structures, and the full theory is actually not diffeomorphism invariant, even though the field equations are defined in terms of tensors. General Relativity in this sense is an exception, for which diffeomorphism invariance at the level of the action and at the level of the field equations yields the same equation: the contracted Bianchi identity. As we have shown, this is not correct for generic theories of gravity since the lack of an action formulation can make the theory nonviable. 

\begin{acknowledgments}
The authors thank M. Gurses for useful discussions. 
E. Altas is supported by TUBITAK Grant No. 123F353.
\end{acknowledgments}

\end{document}